\documentclass[11pt]{article}
\usepackage{amssymb}
\usepackage{epsfig}
\usepackage{graphicx}
\usepackage[nomarkers]{endfloat}
\usepackage{pstricks}

\newcommand{\BE}{\begin{equation}}
\newcommand{\EE}{\end{equation}}
\begin{document}
\begin{titlepage}

\vspace*{1mm}
\begin{center}

{\LARGE{\bf Vacuum structure and ether-drift experiments}}

\vspace*{14mm} {\Large  M. Consoli and L. Pappalardo}
\vspace*{4mm}\\
{\large
Istituto Nazionale di Fisica Nucleare, Sezione di Catania \\
Dipartimento di Fisica e Astronomia dell' Universit\`a di Catania \\
Via Santa Sofia 64, 95123 Catania, Italy \\ }
\end{center}
\begin{center}
{\bf Abstract}
\end{center}

In the data of the ether-drift experiments there might be sizeable
fluctuations  superposed on the smooth sinusoidal modulations due to
the Earth's rotation and orbital revolution. These fluctuations
might reflect the stochastic nature of the underlying "quantum
ether" and produce vanishing averages for all vectorial quantities
extracted from a naive Fourier analysis of the data. By comparing
the typical stability limits of the individual optical resonators
with the amplitude of their relative frequency shift, the presently
observed signal, rather than being spurious experimental noise,
might also express fundamental properties of a physical vacuum
similar to a superfluid in a turbulent state of motion. In this
sense, the situation might be similar to the discovery of the CMBR
that was first interpreted as mere instrumental noise.

\end{titlepage}

\section{Introduction}

In principle, the physical vacuum might behave as a medium with a
non-trivial refractive index. In this case, the speed of light in
the vacuum $c_\gamma={{c}\over{{\cal N}_{\rm vacuum}}}$ would not
coincide exactly with the parameter $c$ entering Lorentz
transformations and therefore could propagate isotropically in only
one frame $\Sigma$. In a moving frame $S'$, there would be an
anisotropy of the two-way speed of light $\bar{c}_\gamma(\theta)$
\BE \label{first}
{{\bar{c}_\gamma(\theta)-\bar{c}_\gamma(0)}\over{c}}\sim ~ B{{
V^2}\over{c^2}}\sin^2\theta \EE with \cite{pagano}\BE \label{second}
|B|\sim 3({\cal N}_{\rm vacuum}- 1) \EE $V$ being the speed of $S'$
with respect to $\Sigma$.  This anisotropy could then be measured
through the beat frequency $\Delta \nu$ of two orthogonal
cavity-stabilized lasers.

In this context, the search for time modulations of the signal
induced by the Earth's rotation (and its orbital revolution) has
always represented a crucial ingredient for the analysis of the
data. For instance, let us consider the relative frequency shift of
two optical resonators for the experiment of Ref.\cite{peters} \BE
\label{basic2}
      {{\Delta \nu (t)}\over{\nu_0}} =
      {S}(t)\sin 2\omega_{\rm rot}t +
      {C}(t)\cos 2\omega_{\rm rot}t
\EE
where $\omega_{\rm rot}$ is the rotation frequency of one resonator
with respect to the other which is kept fixed in the laboratory and
oriented north-south. If one assumes that, for short-time
observations of 2-3 days, the time dependence of a hypothetical
physical signal can only be due to (the variations of the projection
of ${\bf V}$ in the interferometer's plane caused by) the Earth's
rotation, $S(t)$ and $C(t)$ admit the simplest Fourier expansion
($\tau=\omega_{\rm sid}t$ is the sidereal time of the observation in
degrees) \cite{peters} \BE \label{amorse1}
      {S}(t) = S_0 +
      {S}_{s1}\sin\tau +{S}_{c1} \cos\tau
       + {S}_{s2}\sin(2\tau) +{S}_{c2} \cos(2\tau)
\EE \BE \label{amorse2}
      {C}(t) = {C}_0 +
      {C}_{s1}\sin\tau +{C}_{c1} \cos\tau
       + {C}_{s2}\sin(2\tau) +{C}_{c2} \cos(2\tau)
\EE
with time-independent $C_k$ and $S_k$ coefficients. Therefore, by
accepting this theoretical framework, it becomes natural to average
the various $C_k$ and $S_k$ over any 2-3 day period. By further
performing inter-session averages over many short-period
experimental sessions (see Fig.2 of Ref.\cite{joint}) the general
conclusion \cite{joint} is that, although the typical instantaneous
signal is ${\cal O}(10^{-15})$ or larger, the average $C_k$ and
$S_k$ coefficients are at the level ${\cal O}(10^{-17})$ and, with
them, the derived parameters entering the SME \cite{sme} and RMS
\cite{rms} models.

However, as we shall argue in Sect.2, in an alternative picture of
the vacuum, one could consider different types of ether-drift. In
this case, the same experimental data could admit a completely
different interpretation. This possibility will be illustrated in
Sects.3 and 4 where, by comparing with the typical stability limits
of the individual optical resonators, we shall argue that the
presently observed signal is unlike to represent just spurious
instrumental noise. For this reason, in Sect.5, we shall discuss a
possible theoretical framework that can explain the present data. In
this sense, as discussed in the final Sect.6, the situation with the
ether-drift experiments might resemble the discovery of the CMBR
that, at the beginning, was also interpreted as mere instrumental
noise.

\section{The vacuum as a turbulent superfluid}

By accepting the idea that there might be a preferred reference
frame, which is the modern denomination of the old ether, one could
also change perspective and, before assuming any definite
theoretical model, first ask: if light were really propagating in a
physical medium, an ether, and not in a trivial empty vacuum, how
should the motion of (or in) this medium be described ? Namely,
could this relative motion exhibit variations that are {\it not}
only due to known effects as the Earth's rotation and orbital
revolution ?

Without fully understanding the nature of that substratum that we
call physical vacuum, it is not possible to make definite
predictions. Still, according to the standard model of electroweak
and strong interactions, this physical vacuum could be considered as
a superfluid medium \cite{volo} filled by those particle condensates
that play a crucial role for many fundamental phenomena such as mass
generation and quark confinement. By further considering the idea of
a non-zero vacuum energy, this physical substratum could also
represent a preferred reference frame \cite{vacuum}. In this
picture, the standard assumption of smooth sinusoidal variations of
the signal, associated with the Earth's rotation and its orbital
revolution, corresponds to describe the superfluid motion in terms
of simple regular flows.

However, visualization techniques that record the flow of superfluid
helium show \cite{zhang} the formation of macroscopic turbulent
structures with a velocity field that fluctuates randomly around
some average value. The same idea of turbulence, that dates back to
the original Lord Kelvin's attempts to generate transverse waves in
the ether \cite{whittaker}, is also suggested by the formal
equivalence that one can establish \cite{troshkin,tsankov} between
the propagation of small disturbances in a turbulent fluid and the
propagation of electromagnetic waves as described by Maxwell
equations.

To exploit the possible implications for the ether-drift
experiments, let us first recall the general aspects of a turbulent
flow. This is characterized by extremely irregular variations of the
velocity, with time at each point and between different points at
the same instant, due to the formation of eddies \cite{landau}. For
this reason, the velocity continually fluctuates about some mean
value and the amplitude of these variations is {\it not} small in
comparison with the mean velocity itself. The time dependence of a
typical turbulent velocity field can be expressed as \cite{landau}
\BE {\bf v}(x,y,z,t)=\sum_{p_1p_2..p_n} {\bf
a}_{p_1p_2..p_n}(x,y,z)\exp(-i\sum^{n}_{j=1}p_j\phi_j) \EE where the
quantities $\phi_j=\omega_j t+ \beta_j $ vary with time according to
fundamental frequencies $\omega_j$ and depend on some initial phases
$\beta_j$. As the Reynolds number ${\cal R}$ increases, the total
number $n$ of $\omega_j$ and $\beta_j$ increases. In the ${\cal R}
\to \infty$ limit, their number diverges so that the theory of such
a turbulent flow must be a statistical theory.

As discussed by Feynman \cite{feynman}, the idea that there might be
a rich structure of turbulent eddies for very large Reynolds numbers
is, at first sight, quite surprising. After all, starting from the
Navier-Stokes equation, the eddies are generated, in the equation
for the vorticity ${\bf \Omega}=\nabla \times {\bf v}$, by the
viscous term which formally vanishes in the ${\cal R} \to \infty$
limit. However, there is a subtlety. The right-hand term of the
equation \BE {{\partial {\bf \Omega} }\over{\partial t}} + \nabla
\times ({\bf \Omega}\times{\bf v})= {{ 1}\over{{\cal R}}} \Delta{\bf
\Omega} \EE has ${{ 1}\over{{\cal R}}}$ times a second derivative.
This is the higher derivative term in the equation. Thus, although
${{ 1}\over{{\cal R}}}$ becomes smaller and smaller, there can be
rapid variations of ${\bf \Omega}$, producing larger and larger
$\Delta{\bf \Omega}$, that compensate for the small coefficient. For
this reason, the solutions of the viscous equation do {\it not}
approach the solutions of the equation
 \BE {{\partial {\bf
\Omega} }\over{\partial t}} + \nabla \times ({\bf \Omega}\times{\bf
v})= 0 \EE when  ${{ 1}\over{{\cal R}}} \to 0$.

Now, due to the presumably vanishingly small viscosity of a
superfluid ether, the relevant Reynolds numbers are likely
infinitely large in most regimes and we might be faced precisely
with such limit of the theory where the physical vacuum behaves as a
{\it stochastic} medium. In this case random fluctuations of the
physical signal, superposed on the smooth sinusoidal behaviour
associated with the Earth's rotation (and orbital revolution), would
produce deviations of $S(t)$ and $C(t)$ from the simple structure in
Eqs.(\ref{amorse1}) and (\ref{amorse2}) and an effective temporal
dependence of the fitted $C_k=C_k(t)$ and $S_k=S_k(t)$ vectorial
coefficients. In this situation one could easily get, due to phase
interference, vanishing average values $\langle C_k\rangle =\langle
S_k\rangle= 0$.

Nevertheless, as it happens with random fluctuations, the average
{\it amplitude} of the signal could still be preserved. Namely, by
extracting from the data the positive-definite combination \BE A(t)=
\sqrt{S^2(t) +C^2(t)} \EE a definite non-zero $\langle A\rangle$
might well coexist with $\langle C_k\rangle =\langle S_k\rangle= 0$.

\section{Noise or stochastic turbulence ?}

To provide some evidence that indeed we might be faced with this
type of situation, let us consider the experimental apparatus of
Ref.\cite{crossed} where the two optical cavities were obtained from
the same monolithic block of ULE. Due to sophisticated electronics
and temperature controls, the stability limits for the individual
optical cavities are extremely high. More precisely, the effect of
residual amplitude modulation is below 0.02 Hz for both cavities and
thus about $7\cdot 10^{-17}$ for a laser frequency $2.82\cdot
10^{14}$ Hz. For the non-rotating set up, and with power
stabilization, the laser power fluctuations are below $3\cdot
10^{-17}$ for cavity 1 and below $1\cdot 10^{-17}$ for cavity 2.
Finally, for the non-rotating set up, also the tilt instabilities
are below $1\cdot 10^{-16}$, or 0.03 Hz, for both cavities. Thus, by
adding all effects, one deduces a stability of about $\pm 0.05$ Hz
for the individual cavities. This is of the same order of the {\it
mean} frequency shift between the two resonators, say
$\langle{\Delta \nu}\rangle \sim \pm 0.06$ Hz, when averaging the
signal over long temporal sequences.

However, the {\it instantaneous} ${\Delta \nu}$ is much larger, say
$\pm 1$ Hz, and so far has been interpreted as spurious instrumental
noise. To check this interpretation, we observe that, in the absence
of any light anisotropy, the noise in the beat frequency should be
comparable to the noise of the individual resonators. Instead, for
the same non-rotating set up, the plateau of the Allan variance for
the beat signal was found 10 times bigger, namely $1.9\cdot
10^{-15}$ with a corresponding typical shift of about 0.56 Hz (see
Fig.8 of Ref.\cite{crossed}). Also the slopes are different in the
two cases, indicating noises of different nature. The authors tend
to interpret this as cavity thermal noise and refer to
\cite{numata}. However, if the theoretical estimate of
Ref.\cite{numata} applies, the relevant effect should be
considerably smaller, about $4\cdot 10^{-16}$ or 0.13 Hz (see the
first entry in Table I of \cite{numata}). In any case, one can also
compare with experiments performed in the cryogenic regime. If this
typical ${\cal O}(10^{-15})$ beat signal reflects the stochastic
nature of an underlying quantum ether, it should be found in these
different experiments as well.

\section{An alternative analysis of the data}

Motivated by the previous arguments, we have explored the idea that
the observed beat signal could be due to some form of turbulent
ether flow. In principle, in this perspective, one should abandon
the previous type of analysis based on a fixed preferred reference
frame and extract the average amplitude from the instantaneous data
before any averaging procedure. However, these instantaneous data
are not available and thus, to extract $\langle A\rangle$, that was
never reported by the experimental groups, one can only use the
$C_k$ and $S_k$ coefficients averaged within each 2-3 day session.
Due to the expected negative phase interference, the resulting
$\langle A\rangle$ should represent a lower limit for its true
experimental value.

To extract $\langle A\rangle$,  one should first re-write
Eq.(\ref{basic2}) as \BE \label{basic3}
      {{\Delta \nu (t)}\over{\nu_0}} =
      A(t)\cos (2\omega_{\rm rot}t -2\theta_0(t))
\EE with \BE \label{interms}
C(t)=A(t)\cos2\theta_0(t)~~~~~~~~S(t)=A(t)\sin2\theta_0(t)\EE
$\theta_0(t)$ representing the instantaneous direction of the
ether-drift effect in the plane of the interferometer. In this
plane, the projection of the full ${\bf V}$ is specified by its
magnitude $v=v(t)$ and by its direction $\theta_0=\theta_0(t)$
(counted by convention from North through East so that North is
$\theta_0=0$ and East is $\theta_0=\pi/2$). If one assumes
Eqs.(\ref{amorse1}) and (\ref{amorse2}), then $v(t)$ and
$\theta_0(t)$ can be obtained from the relations
\cite{nassau,dedicated} \BE \label{cosine}
       \cos z(t)= \sin\gamma\sin \phi + \cos\gamma
       \cos\phi \cos(\tau-\alpha)
\EE \BE
       \sin z(t)\cos\theta_0(t)= \sin\gamma\cos \phi -\cos\gamma
       \sin\phi \cos(\tau-\alpha)
\EE \BE
       \sin z(t)\sin\theta_0(t)= \cos\gamma\sin(\tau-\alpha) \EE
\BE \label{projection}
       v(t)=V \sin z(t) ,
\EE where $\alpha$ and $\gamma$ are respectively the right ascension
and angular declination of ${\bf{V}}$. Further, $\phi$ is the
latitude of the laboratory and $z=z(t)$ is the zenithal distance of
${\bf{V}}$. Namely, $z=0$ corresponds to a ${\bf{V}}$ which is
perpendicular to the plane of the interferometer and $z=\pi/2$ to a
${\bf{V}}$ that lies entirely in that plane. From the above
relations, by using the ${\cal O}(v^2/c^2)$ relation $A(t) \sim
{{v^2(t)}\over{c^2}}$, the other two amplitudes $S(t)=A(t)\sin
2\theta_0(t)$ and $C(t)=A(t)\cos 2\theta_0(t)$ can be obtained up to
an overall proportionality constant. By using the expressions for
$S(t)$ and $C(t)$ reported in Table I of Ref.~\cite{peters} (in the
RMS formalism \cite{rms}), this proportionality constant turns out
to be ${{1}\over{2}}|B|$  so that we finally find \BE
\label{amplitude1}
       A(t)= {{1}\over{2}}|B| {{v^2(t) }\over{c^2}}
\EE where $B$ is the anisotropy parameter entering the two-way speed
of light Eq.(\ref{first}). It is a simple exercise to check that, by
using Eqs.(\ref{interms}), Eqs.(\ref{cosine})-(\ref{amplitude1}) and
finally replacing $\chi=90^o-\phi$, one re-obtains the expansions
for $C(t)$ and $S(t)$ reported in Table I of Ref.~\cite{peters}.

We can then replace Eq.~(\ref{projection}) into
Eq.~(\ref{amplitude1}) and, by adopting a notation of the type in
Eqs.(\ref{amorse1})-(\ref{amorse2}), express the Fourier expansion
of $A(t)$ as
\BE \label{amorse}
       A(t) = A_0 +
       A_1\sin\tau +A_2 \cos\tau
        +  A_3\sin(2\tau) +A_4 \cos(2\tau)
\EE
where (the daily averaging of any quantity is here denoted by
$\langle..\rangle$),
\BE \label{aa0}
   \langle
A\rangle  =   A_0 ={{1}\over{2}}|B| {{\langle
v^2(t)\rangle}\over{c^2}}=
       {{1}\over{2}}|B| {{V^2}\over{c^2}}
       \left(1- \sin^2\gamma\cos^2\chi
       - {{1}\over{2}} \cos^2\gamma\sin^2\chi \right)
\EE
\BE \label{a1}
       A_1=-{{1}\over{4}}|B| {{V^2}\over{c^2}}\sin 2\gamma
       \sin\alpha \sin 2\chi
~~~~~~~~~~~~~~~
       A_2=-{{1}\over{4}}|B| {{V^2}\over{c^2}}\sin 2\gamma
       \cos\alpha \sin 2\chi
\EE
\BE \label{a3}
       A_3=-{{1}\over{4}}  |B| {{V^2}\over{c^2}}\cos^2 \gamma
       \sin 2\alpha \sin^2 \chi
~~~~~~~~~~~~~~~
       A_4=-{{1}\over{4}} |B| {{V^2}\over{c^2}} \cos^2 \gamma
       \cos 2\alpha \sin^2 \chi
\EE To obtain $A_0$ from the $C_k$ and $S_k$, we observe that by
using Eq.(\ref{amorse}) one obtains \BE \label{amplitude0}\langle
~A^2(t)~ \rangle= A^2_0+
{{1}\over{2}}(A^2_{1}+A^2_{2}+A^2_{3}+A^2_{4})\EE On the other hand,
by using Eqs.(\ref{amorse1}), (\ref{amorse2}) and (\ref{interms}),
one also obtains
 \BE
\label{amplitude}\langle ~A^2(t)~ \rangle= \langle ~C^2(t) +S^2(t)~
\rangle=C^2_0 + S^2_0 + Q^2 \EE where \BE \label{Q} Q= \sqrt{
{{1}\over{2}}(C^2_{11}+S^2_{11}+C^2_{22}+S^2_{22})} \EE and  \BE
\label{csid}
      {C}_{11}\equiv \sqrt{{C}^2_{s1}
      + {C}^2_{c1}}
~~~~~~~~~~~~~~~~
      {C}_{22}\equiv \sqrt{{C}^2_{s2}
      + {C}^2_{c2}}
\EE
 \BE \label{s2sid}
      {S}_{11}\equiv \sqrt{{S}^2_{s1}
      + {S}^2_{c1}}
~~~~~~~~~~~~~~~~
 {S}_{22}\equiv \sqrt{{S}^2_{s2}
      + {S}^2_{c2}}
\EE
Therefore, one can combine the two relations and get \BE
\label{final} A^2_0(1+r)= C^2_0 + S^2_0 + Q^2 \EE where
 \BE r\equiv
{{1}\over{2A^2_0}}(A^2_{1}+A^2_{2}+A^2_{3}+A^2_{4}) \EE By computing
the ratio $r=r(\gamma,\chi)$ with Eqs.(\ref{aa0})-(\ref{a3}), one
finds \BE \label{range} 0\leq r\leq 0.4\EE for the latitude of the
laboratories in Berlin \cite{peters} and D\"usseldorf
\cite{schiller} in the full range $0 \leq |\gamma|\leq \pi/2$.  We
can thus define an average amplitude, say $\hat{A}_0$, which is
determined in terms of $Q$ alone as \BE \label{AQ} \hat{A}_0=
{{Q}\over{\sqrt{1+r}}} \sim (0.92 \pm 0.08)Q \EE where the
uncertainty takes into account the numerical range of $r$ in
Eq.(\ref{range}). This quantity provides, in any case, a {\it lower
bound} for the true experimental $A_0$ since \BE \label{lower}
A_0=\sqrt{ {{C^2_0 + S^2_0 +Q^2}\over{1+r}}
 } \geq \hat{A}_0 \EE
At the same time $Q$ is determined only by the $C_{s1}$,
$C_{c1}$,... and their S-counterparts. According to the authors of
Refs.\cite{peters,joint}, these coefficients are much less affected
by spurious effects, as compared to $C_0$ and $S_0$, and so will be
our amplitude $\hat{A}_0$.

\begin{table}
\caption{By using Eqs. (\ref{Q}), (\ref{csid}) and (\ref{s2sid}), we
report the values $Q_i$, their uncertainties $\Delta Q_i$ and the
ratio $R_i=Q_i/\Delta Q_i$ for each of the 27 experimental sessions
of Ref.\cite{joint}. These values have been extracted, according to
standard error propagation for a composite observable, from the
basic $C_k$ and $S_k\equiv B_k$ coefficients reported in Fig.2 of
Ref.\cite{joint}.}
\begin{center}
\label{tab:3}
\begin{tabular}{lll}
\hline\noalign{\smallskip}
 $ Q_i  [{\rm x}10^{-16}]$ & $ \Delta Q_i  [{\rm
x}10^{-16}]$ & $R_i= Q_i/\Delta Q_i  $  \\
\noalign{\smallskip}\hline\noalign{\smallskip} \hline
$13.3$ & ~~~ $3.4$ &~~~  $3.9$  \\
$14.6$ & ~~~ $4.8$ & ~~~ $3.0$  \\
$6.6$ &  ~~~ $2.6$ & ~~~ $2.5$  \\
$17.8$ & ~~~ $2.8$ & ~~~ $6.3$ \\
$14.0$ & ~~~ $5.8$ & ~~~ $2.5$ \\
$11.1$ & ~~~ $4.2$ & ~~~ $2.6$ \\
$13.0$ & ~~~ $4.2$ & ~~~ $3.1$ \\
$19.2$ & ~~~ $6.1$ & ~~~ $3.1$ \\
$13.0$ & ~~~ $4.7$ & ~~~ $2.8$ \\
$12.0$ & ~~~ $3.5$ & ~~~ $3.4$ \\
$5.7$ &  ~~~ $2.4$ &  ~~~ $2.4$ \\
$14.6$ &  ~~~ $5.2$ & ~~~ $2.8$ \\
$16.9$ &  ~~~ $3.3$ & ~~~ $5.1$ \\
$8.3$ &  ~~~ $2.4$ &  ~~~ $3.4$ \\
$27.7$ & ~~~ $4.5$ & ~~~ $6.2$  \\
$28.3$ & ~~~ $5.7$ & ~~~ $5.0$  \\
$12.7$ & ~~~ $2.5$ & ~~~ $5.1$ \\
$12.1$ & ~~~ $5.3$ & ~~~ $2.3$ \\
$13.7$ & ~~~ $6.0$ & ~~~ $2.3$ \\
$23.9$ & ~~~ $5.7$ & ~~~ $4.2$ \\
$28.9$ & ~~~ $4.3$ & ~~~ $6.7$ \\
$18.4$ & ~~~ $5.1$ & ~~~ $3.6$ \\
$19.2$ & ~~~ $6.2$ & ~~~ $3.1$ \\
$11.9$ & ~~~ $2.7$ & ~~~ $4.4$ \\
$18.1$ & ~~~ $5.4$ & ~~~ $3.3$ \\
$4.2$ & ~~~  $2.9$ &  ~~~ $1.4$ \\
$31.6$ &~~~  $7.9$ & ~~~ $4.0$ \\
\noalign{\smallskip}\hline
\end{tabular}
\end{center}
\end{table}

We have computed the $Q$ values for the 27 short-period experimental
sessions of Ref.\cite{joint}. Their values are reported in Table I.
Although more recent data were announced in Ref.\cite{sme}, only the
results of fits within the SME model, and not the data themselves,
are explicitly reported. In any case, the data of Ref.\cite{joint}
represent, within their statistics, a sufficient basis to deduce
that a rather stable pattern is obtained. This is due to the
rotational invariant character of $Q$ in the 8-th dimensional space
of the $C_{s1}$, $C_{c1}$,...$S_{s2}$, $S_{c2}$ so that variations
of the individual coefficients tend to compensate. By taking an
average of these 27 determinations one finds a mean value \BE
\label{Qmean1} \bar{Q}=\left(13.0 \pm 0.7 \pm 3.8\right)\cdot
10^{-16} ~~~~~~~~~~~~{\rm Ref.\cite{joint}}\EE where the former
error is purely statistical and the latter represents an estimate of
the systematical effects. Such systematic uncertainty, deduced from
the data of Ref.\cite{joint}, is also completely consistent with the
comparable stability limit $4\cdot 10^{-16}$ of reference optical
cavities \cite{nist} for integration times ${\cal O}(10^2)$ seconds
as those appropriate for the basic data of the Berlin experiment
(collected with sets of 10 turntable rotations of 43 seconds each).

As anticipated, for a further control of the validity of our
analysis, we have compared with the cryogenic experiment of
Ref.\cite{schiller}. In this case, we have obtained the analogous
value \BE Q=(13.1 \pm 2.1)\cdot 10^{-16}~~~~~~~~~~~~~~~{\rm
Ref.\cite{schiller}} \EE  from the corresponding $C_k$ and $S_k$
coefficients. Thus, by using Eq.(\ref{AQ}) and the two values of $Q$
reported above, we obtain \BE \label{final1} \hat{A}^{\rm
exp}_0=(12.0 \pm 1.0 \pm 3.5)\cdot 10^{-16}~~~~~~~~~~~~{\rm
Ref.\cite{joint}} \EE \BE \label{mean2} \hat{A}^{\rm exp}_0= (12.1
\pm 1.0 \pm 2.1)\cdot 10^{-16}~~~~~~~~~~~~~~{\rm
Ref.\cite{schiller}}\EE where the former uncertainty takes into
account the variation of $r$ in Eq.(\ref{range}) and the latter is
both statistical and systematical.

We emphasize that this stable value of about $12\cdot 10^{-16}$ is a
lower limit for the true experimental $A_0$. At the same time, it
could hardly be interpreted as a spurious effect of experimental
noise. In fact,
 from the estimates of Ref.\cite{numata}, based on the
fluctuation-dissipation theorem, there is no reason to expect that
experiments running in so different conditions should exhibit the
same experimental noise.

\section{A possible theoretical framework}

Let us now explore a possible theoretical framework that can explain
the above experimental values. To this end, we shall first use
Eq.(\ref{aa0}) and Eq.(\ref{second}) and re-write the theoretical
prediction as \BE \label{theory3} A^{\rm th}_0 = {{3}\over{2}}({\cal
N}_{\rm vacuum}- 1) {{V^2}\over{c^2}} f(\gamma,\chi) \EE with  \BE
f(\gamma,\chi)=
       \left(1- \sin^2\gamma\cos^2\chi
       - {{1}\over{2}} \cos^2\gamma\sin^2\chi \right) \EE
Then, in a flat-space picture of gravity, for an apparatus placed on
the Earth's surface the vacuum refractive index was estimated to
have the value \cite{pagano} \BE \label{refractive}{\cal N}_{\rm
vacuum}- 1 \sim {{2GM}\over{c^2R}} \sim 1.4\cdot 10^{-9}\EE $G$
being Newton's constant and $M$ and $R$ the Earth's mass and radius.
Since the meaning of "flat-space picture of gravity" might be
ambiguous, we shall repeat here the basic derivation.

The usual interpretation of phenomena in gravitational fields is in
terms of a fundamentally curved space-time. However, some authors
\cite{wilson, dicke, puthoff} have argued that, as light in
Euclidean space deviates from a straight line in a medium with
variable density, an effective curvature might also be the
consequence of a suitable polarization of the physical flat-space
vacuum. The substantial phenomenological equivalence with the
standard interpretation was well summarized by Atkinson as follows
\cite{atkinson} : "It is possible, on the one hand, to postulate
that the velocity of light is a universal constant, to define {\it
natural} clocks and measuring rods as the standards by which space
and time are to be judged and then to discover from measurement that
space-time is {\it really} non-Euclidean. Alternatively, one can
{\it define} space as Euclidean and time as the same everywhere, and
discover (from exactly the same measurements) how the velocity of
light and natural clocks, rods and particle inertias {\it really}
behave in the neighborhood of large masses." In this sense, in the
flat-space approach, one is adopting a "Lorentzian perspective"
where physical rods and clocks are held together by the same forces
underlying the structure of the "ether" (the physical vacuum). Thus
the Equivalence Principle actually means that the effect of an
external gravitational field can be re-absorbed into the space-time
units of a freely-falling observer so as to preserve local Lorentz
covariance.

Notice that the Equivalence Principle was introduced \cite{pre}
before General Relativity and, as such, does not rely on the notion
of a fundamentally curved space-time. Actually, it could even be
considered misleading within a pure, curved space-time context
described by a given Riemann tensor \cite{synge}. For this reason,
it has also been interpreted in terms of an ether flow
\cite{kirkwood}. Nevertheless, regardless of its ultimate physical
origin, this fundamental principle implies that in a freely-falling
frame, given two space-time events that differ by $(dx,dy,dz,dt)$
and the local space-time metric \BE ds^2=c^2dt^2-
(dx^2+dy^2+dz^2)\EE one gets from $ds^2=0$ the same speed of light
as in the absence of any gravitational effect.

However, to a closer look, an observer placed on the Earth's surface
is equivalent to a freely-falling frame {\it up to the presence of
the Earth's gravitational field}. This leads to the weak-field,
isotropic modifications of the metric \BE \label{iso} ds^2=
c^2dt^2(1-{{2GM}\over{c^2R}})-(1+{{2GM}\over{c^2R}})
(dx^2+dy^2+dz^2)=c^2dt^2_0-dl_0^2\EE Here $dt_0$ and $dl_0$ denote
respectively the elements of proper time and proper length in terms
of which, in General Relativity, one would again deduce from
$ds^2=0$ the same universal value ${{dl_0}\over{dt_0}}=c$. On the
other hand, in the flat-space approach the condition $ds^2=0$ is now
interpreted in terms of a refractive index for the vacuum as in
Eq.(\ref{refractive}) and thus light can be seen isotropic in {\it
only one} reference frame, say $\Sigma$. This means that isotropy is
only valid if the Earth were at rest in the hypothetical $\Sigma$.
Otherwise, by introducing the Earth's velocity components $V_i$,
off-diagonal elements $g_{0i}\sim V_i/c$ in the effective metric are
generated by the Lorentz transformation for the observer placed on
the Earth. These non-zero $g_{0i}$ can be imagined as being due to a
directional polarization of the vacuum induced by the now moving
Earth's gravitational field and express the general property
\cite{volkov} that any metric, locally, can be brought into diagonal
form by suitable rotations and boosts. By starting in $\Sigma$ with
the diagonal and isotropic form Eq.(\ref{iso}), one obtains \BE
g_{0i} \sim 2 ({\cal N}_{\rm vacuum}- 1)~{{V_i}\over{c}} \EE and an
anisotropy of the speed of light. The value of light anisotropy can
also be computed by simply composing the isotropic speed $c/{\cal
N}_{\rm vacuum}$  with a given velocity ${\bf V}$. In this way, one
predicts a two-way speed of light as in Eqs.(\ref{first}) and
(\ref{second}). \vskip 10 pt

\section{Conclusions}

Let us now collect all previous results. For an average colatitude
$\chi\sim$ 38 degrees (as for the average location of the Berlin and
D\"usseldorf laboratories) one finds $f(\gamma,\chi)=(0.60 \pm
0.22)$ in the whole range $0 \leq |\gamma| \leq \pi/2$. Therefore,
by using (\ref{refractive}) in Eq.(\ref{theory3}), one finds \BE
\label{theory4} A^{\rm th}_0 = (12.6\pm 4.6)\cdot 10^{-16}~(
V/300~{\rm km \cdot s^{-1}})^2 \EE in units of the typical speed 300
km/s of most cosmic motions. This estimate is equivalent to the
previous prediction \cite{pla} for the amplitude of the frequency
shift measured in a symmetrical apparatus (i.e. with two orthogonal
rotating resonators) \BE A^{\rm symm}_0=|B|{{\langle
v^2\rangle}\over{c^2}}=2A^{\rm th}_0 =(1.9 \pm 0.7)\cdot 10^{-15}
\EE in terms of the average projected speed $\sqrt{\langle v^2
\rangle}$=$(204\pm 36)$ km/s obtained from a re-analysis of the
classical experiments.

Since these theoretical estimates agree well with the experimental
values (\ref{final1})-(\ref{mean2}) extracted from
Refs.\cite{joint,schiller}, and with the mean amplitude $1.9 \cdot
10^{-15}$ of the signal measured in the symmetric apparatus of
Ref.\cite{crossed}, we conclude that the observed frequency shift,
rather than being spurious noise of the underlying optical cavities,
might also reflect basic properties of the vacuum. On the one hand,
this appears as a polarizable medium responsible for the apparent
curvature effects induced by a gravitational field. On the other
hand, the observed strong random fluctuations of the signal support
the view of a stochastic medium, similar to a superfluid in a
turbulent state of motion. In this sense, the situation might be
similar to the discovery of the CMBR that, at the beginning, was
also interpreted as mere instrumental noise.

A crucial test of our interpretation can be performed in a
freely-falling spacecraft. In this case, where the vacuum refractive
index ${\cal N}_{\rm vacuum}$ for the freely-falling observer is
exactly unity, the typical instantaneous $\Delta \nu$ should be much
smaller (by orders of magnitude) than the corresponding  ${\cal
O}(10^{-15})$ value measured with the same interferometer on the
Earth's surface.

\end{document}